\documentclass[11pt]{article}

\renewcommand{\vec}[1]{\mbox{\boldmath ${#1}$}}
\newcommand{\U}{{\rm U}}
\newcommand{\half}{\mbox{$\f{1}{2}$}}
\newcommand{\f}{\frac}

\newcommand{\eq}{\begin{equation}}
\newcommand{\eqend}{\end{equation}}
\newcommand{\eqa}{\begin{eqnarray}}
\newcommand{\nonueqa}{\begin{eqnarray*}}
\newcommand{\eqaend}{\end{eqnarray}}
\newcommand{\nonueqaend}{\end{eqnarray*}}
\newcommand{\nonu}{\nonumber \\ \nopagebreak}
\newcommand{\bma}[1]{\begin{array}{#1}}
\newcommand{\ema}{\end{array}}
\newcommand{\bc}{\begin{center}}
\newcommand{\ec}{\end{center}}

\newcommand{\Ref}[1]{(\ref{#1})}

\newcommand{\dd}{{\rm d}}
\newcommand{\ee}[1]{\mbox{{\rm e}}^{#1}}
\newcommand{\ii}{{\rm i}}

\newcommand{\om}{\omega}
\renewcommand{\phi}{\varphi}
\newcommand{\del}{\delta}

\newfam\msbfam
\batchmode\font\twelvemsb=msbm10 scaled\magstep1 \errorstopmode
\ifx\twelvemsb\nullfont\def\Bbb{\bf}
\else   \catcode`\@=11
        \font\tenmsb=msbm10 \font\sevenmsb=msbm7 \font\fivemsb=msbm5
        \textfont\msbfam=\tenmsb
        \scriptfont\msbfam=\sevenmsb \scriptscriptfont\msbfam=\fivemsb
        \def\Bbb{\relax\ifmmode\expandafter\Bbb@\else
                \expandafter\nonmatherr@\expandafter\Bbb\fi}
        \def\Bbb@#1{{\Bbb@@{#1}}}
        \def\Bbb@@#1{\fam\msbfam\relax#1}
        \catcode`\@=\active
\fi
\newcommand{\R}{{\Bbb R}}

\newcommand{\Z}{{\Bbb Z}}

\newcommand{\cC}{{\cal C}}
\newcommand{\cL}{{\cal L}}
\newcommand{\cH}{{\cal H}}
\newcommand{\cE}{{\cal E}}
\newcommand{\cG}{{\cal G}}
\newcommand{\cU}{{\cal U}}

\newcommand{\ccr}[2]{{[} {#1},{#2} {]} }
\newcommand{\car}[2]{{\{} {#1},{#2} {\}} }
\newcommand{\normal}[1]{\, :\! {#1}\! :\, \;}

\renewcommand{\vec}[1]{{\bf {#1} }}

\newcommand{\rk}{{\rm rank}}
\newcommand{\ia}{{\rm A}}
\newcommand{\iv}{{\rm V}}

\newcommand{\vc}{\vec{c}}
\newcommand{\vC}{\vec{C}}
\newcommand{\mOm}{{\bf h }}
\newcommand{\mom}{{\bf D}}
\newcommand{\mU}{{\bf U}}
\newcommand{\mM}{{\bf Y}}

\newcommand{\mV}{{\bf V}}
\newcommand{\mUt}{{\bf {\tilde U} }}
\newcommand{\mq}{{\bf q}}
\newcommand{\vn}{\vec{n}}
\newcommand{\vf}{\vec{f}}
\newcommand{\vnt}{\vec{n}^T}
\newcommand{\vm}{\vec{m}}

\newcommand{\vk}{\vec{k}}
\newcommand{\ve}{\vec{e}}
\newcommand{\vet}{\vec{e}^T}
\newcommand{\vE}{\vec{E}}
\newcommand{\vEt}{\vec{E}^T}
\newcommand{\vF}{\vec{F}}
\newcommand{\vFt}{\vec{F}^T}

\newcommand{\M}{M}
\newcommand{\RR}{{\bf R}}

\newcommand{\id}{{\bf 1}}
\newcommand{\Nf}{N_{\rm F}}
\newcommand{\Np}{N_{\rm P}}

\newcommand{\dLam}{\Lambda^*}
\newcommand{\dLamF}{\Lambda^*_{odd}}

\newcommand{\dInt}{\hat{\int}_{\dLam}}
\newcommand{\dIntF}{\hat{\int}_{\dLamF}}

\newcommand{\Int}{\int_{-L/2}^{L/2}}
\newcommand{\ddel}{\hat{\delta}}
\newcommand{\Del}{\hat{\rm d}}

\begin{document}

\begin{flushright}
April 25, 2000
\end{flushright}
\vspace{1.5cm}
\renewcommand{\thefootnote}{\alph{footnote}}
\begin{center}

{\Large \bf Chiral Schwinger models without gauge anomalies}\\

\vspace{1.5cm}
{\large Harald Grosse$^*$ and Edwin Langmann$^{**}$} \\

\vspace*{0.5cm}
\noindent
$^{*,**}${\em Erwin Schr\"odinger Internat.\ Inst.\ for
Mathematical Physics, A-1090 Wien, Austria}\\
$^*${\em Institut f\"ur Theoretische Physik, 
Universit\"at Wien, A-1090 Wien, Austria} \\
$^{**}${\em Theoretical Physics, Royal Institute of Technology, 
S-100 44 Stockholm, Sweden}\\
\vspace{2cm}
\end{center}

\setcounter{footnote}{0}
\renewcommand{\thefootnote}{\arabic{footnote}}

\begin{abstract}
We find a large class of quantum gauge models with 
massless fermions where the coupling to the gauge fields is not
chirally symmetric and which nevertheless do not suffer from 
gauge anomalies. To be specific we study two dimensional Abelian models 
in the Hamiltonian framework which can be
constructed and solved by standard techniques. The general model
describes $\Np$ photon fields and $\Nf$ flavors of Dirac fermions 
with  $2\Nf\Np$ different coupling constants i.e.\ the chiral 
component of each fermion can be coupled to the gauge fields differently. 
We construct these models and find conditions so that no gauge 
anomaly appears. If these conditions hold 
it is possible to construct and solve the model explicitly,  
so that gauge- and Lorentz invariance are manifest. 
\end{abstract}

\subsection*{1. Introduction} 
In this paper we discuss a large class of gauge theory models where
the gauge fields are coupled to $\Nf$ Dirac fermions with different 
coupling constants (= charges) $e_{j,\sigma}^\alpha$ where the index 
$\sigma=+,-$ distinguishes the chiral components of the fermions, 
$j=1,2,\ldots, \Nf$ is a fermion flavor index, and $\alpha=
1,2,\ldots\Np$ labels different gauge fields. As we shall see, 
the existence of large gauge transformations implies that the 
charges have to be quantized i.e.\ $e_{j,\sigma}^\alpha = 
n_{j,\sigma}^\alpha e^\alpha$ (no summation) with 
$n_{j,\sigma}^\alpha$ integers. 
We say that such a model is {\em chirally symmetric} if
for all $\alpha$ and $j$, 
$n^\alpha_{j,+}=n^\alpha_{\pi(j),-}$ 
for some permutation $\pi$. 

It is worth noting that the standard Schwinger model
is obtained as the special case $\Np=\Nf=1$ and $n_+=n_-$, and the 
chiral Schwinger model corresponds to  $\Np=\Nf=1$ and $n_+=1$ and $n_-=0$
(for review on previous work on these and similar models see
\cite{NN}). The class of models with only one photon field 
($\Np=1$) and coupling constants $e n_{j,+}$ and $e n_{j,-}$ 
to the right- and left handed chiral components of the fermions, 
respectively ($j=1,\ldots,\Nf$, and the $n_{j,\sigma}$ are 
integers) was previously proposed and studied 
in \cite{NN}. For these models it is known
that the gauge anomaly cancels whenever 
\eq
\label{m345}
\sum_{j=1}^{\Nf} \left( n_{j,+}^2-n_{j,-}^2 \right) = 0 \: . 
\eqend 
The simplest non-trivial 
(i.e.\ chirally asymmetric) solution is for $\Nf=2$, 
$n_{j,+}=(3,4)$ and $n_{j,-}=(0,5)$, and therefore one 
sometimes refers to this model as the {\em 3-4-5 model} 
\cite{NN}. Similarly one can find non-trivial examples for all
Pythagorean Triple i.e.\ integer solutions of 
$a^2+b^2=c^2$. 

For the general class of models with $\Nf$ fermions and 
$\Np\leq \Nf$ photons which we study we find
the following conditions for gauge anomalies to be absent, 
\eq
\label{m345_1}
\sum_{j=1}^{\Nf} \left( n_{j,+}^\alpha n_{j,+}^\beta
-n_{j,-}^\alpha n_{j,-}^\beta  \right) = 0 \qquad 
\forall \alpha,\beta=1,\ldots,\Np\: .  
\eqend 
A simple non-trivial example is for $\Nf=\Np=2$, 
$n^1_{j,+}=(3,4)$, $n^1_{j,-}=(0,5)$,  
$n^2_{j,+}=(-1,2)$, and $n^2_{j,-}=(2,1)$. 
To find solutions of these conditions for given $\Np$ and as small 
$\Nf$ as possible seems to be an interesting generalization of the 
problem of finding Pythagorean Triples.  

The study of the 3-4-5 model in Ref.\ \cite{NN}
was in the path integral formalism. Here we use a Hamiltonian
approach in the spirit of Ref.\ \cite{Manton}. This approach allows a
rigorous construction of these model in the Hamiltonian framework
using the quasi-free representation of boson- and fermion 
field algebras \cite{LS,GLR}, and we also can solve the model using 
standard bosonization techniques \cite{LM}. Our general class of models 
(i.e.\ arbitrary $\Nf$ and $\Np$) is quite complicated, and it 
is somewhat surprising that it is possible to find the solution 
in such an explicit manner as we do in this paper. It is also 
intriguing to see the importance of the no-anomaly conditions 
in  Eq.\ \Ref{m345_1} at several different, seemingly unrelated 
points of our construction and solution. 

To simplify notation we explain our methods and computations in 
detail for the simplest case $\Np=1$, and we are careful to 
do things such that the generalization to the case $\Np>1$
is easy. 

The plan of this paper is as follows. We first concentrate on the
models with one photon field and an arbitrary number $\Nf$ of 
fermion fields. After a formal definition  we summarize 
the rigorous construction of these model in our framework. 
Anomalies are a consequence of Schwinger terms which result 
from the normal ordering necessary to construct the fermion currents 
as well-defined  operators on the Hilbert space of the model.
Especially we compute the gauge anomalies from the commutators of the 
implementers of gauge transformations (= Gauss' law generators), 
and Eq.\ \Ref{m345} is obtained as condition for a vanishing gauge anomaly. 
If and only if this latter condition holds
a simple construction and solution of this model is obtained. 
As mentioned, these results for $\Np=1$ are presented such that 
the generalization to our general class of models with 
$\Np$ photons and $\Nf$ fermions is easy. The presentation of
our results for this latter case are given in a rather short final 
paragraph.

\subsection*{2. Notation} 
Throughout this paper we consider Abelian gauge theories with massless 
fermions on two dimensional spacetime which is a cylinder, 
$\R\times S^1$, parametrized by coordinates $(x^\mu)=(x^0,x^1)$ where 
$x^0=t\in\R$ (= time) and $x^1=x\in[-L/2,L/2]$ (= spatial
coordinate; $0<L<\infty$ is the spatial length). Our metric tensor
is $diag(1,-1)$. We use the Einstein summation convention {\em only}
for spacetime indices $\mu,\nu=0,1$ but {\em not} for flavor- or 
spin indices.

\subsection*{3. Formal definition of the model}
We now define in some detail the simplest non-trivial example for 
a chiral model without gauge anomalies. As described above, 
this model contains $\Nf$ flavors of Dirac fermion fields $\psi_{j,\sigma}$,
$\bar\psi_{j,\sigma}$ and one Abelian gauge field $A_\mu$; here and 
in the following, $\sigma,\sigma' = +,- $ are spin indices, 
and indices $j,j'= 1,\ldots,\Nf$ 
distinguish the different fermion flavors.

The model is formally defined by the Lagrangian 
\eq
\label{2} 
\cL = -\frac{1}{4} F_{\mu\nu}F^{\mu\nu} + 
\sum_{j=1}^{\Nf}\bar\psi_j \gamma^\mu[-\ii\partial_\mu + 
e (n_{j,+}P_+ +n_{j,-}P_-) A_\mu ] 
\psi_j 
\eqend
where $F_{\mu\nu}= \partial_\mu A_\nu-\partial_\nu A_\mu$, 
and
\eq
P_\pm= \half(\id \pm \gamma_3)
\eqend	
are chiral projections; here 
$\gamma^\mu\equiv (\gamma^\mu)_{\sigma,\sigma'}$ are Dirac matrices
which we take as $\gamma^0=\sigma_1$, $\gamma^1=-\ii\sigma_2$, and
$\gamma_3= \gamma^0\gamma^1=\sigma_3$ ($\sigma_{1,2,3}$ are the
Pauli spin matrices as usual), 
and the real parameters 
$e_{j,\sigma}=e n_{j,\sigma}$ are coupling constants.
The gauge group for this model is $\cG=C^\infty(\R\times S^1;\U(1))$
(= smooth $\U(1)$-valued functions on spacetime), and the 
Lagrangian Eq.\ \Ref{2} is obviously 
invariant under the following gauge transformations,
\eqa
\label{gt}
\psi_j &\to& (\ee{\ii n_{j,+}\chi}P_+ +\ee{\ii n_{j,-}\chi}P_-)\psi_j\nonu
\bar\psi_j &\to& \bar\psi_j (\ee{-\ii n_{j,+}\chi}P_-
+\ee{-\ii n_{j,-}\chi}P_+)
   \nonu
A_\mu &\to& A_\mu - \frac{1}{e}\partial_\mu\chi
\eqaend
for all $\ee{\ii\chi}\in \cG$. Note that 
the  existence of the large gauge transformation 
$\ee{\ii\chi(x,t)}=\ee{\ii x/L}$
forces us to require that the $n_{j,\sigma}$ are integers
(otherwise the large gauge transformations cannot be implemented in
our model): The charges of the different fermion flavors have to 
be quantized. 

To motivate our construction of the model in the Hamiltonian framework
below we recall some formulas from the formal canonical procedure; see
e.g.\ \cite{Sundermeyer}. From the action Eq.\ \Ref{2} one 
computes the canonical momenta for the various fields of the model and
obtains the following canonical (anti-) commutator
relations\footnote{$\car{a}{b}=ab+ba$ and
$\ccr{a}{b}=ab-ba$}$^,$\footnote{here and in
the following we set $t=0$ and make explicit the dependence on the
spatial coordinate only} 
\eqa
\label{CACR}
\car{ \psi_{j,\sigma}(x)}{\psi^\dag_{j',\sigma'}(y) } &=&
\del_{\sigma,\sigma'}\del_{j,j'} \del(x-y) 
\nonu
\ccr{ E(x)}{A_1(y)} &=& -\ii\del(x-y)
\eqaend
etc.\
where $\psi^\dag=\bar\psi\gamma^0$ and $E=F_{01}$.
Moreover, with the notation 
\eqa
H_0 &=& \sum_{j=1}^{\Nf}  \Int \dd{x}\,  
\psi_j^\dag(x)\gamma_3(-\ii\partial_1)\psi_j(x)\nonu
J(x)&=&\sum_{j=1}^{\Nf}\sum_{\sigma=\pm} n_{j,\sigma} \psi^\dag_j(x) 
\gamma_3 P_\sigma \psi_j(x)\nonu
\rho(x) &=&\sum_{j=1}^{\Nf} \sum_{\sigma=\pm}
n_{j,\sigma} \psi^\dag_j(x) P_\sigma 
\psi_j(x)  
\eqaend
the resulting Hamiltonian for the model can be written as
\eq
\label{Ham}
H = H_0 + \Int \dd{x} \left( \half E(x)^2 + e A_1(x) J(x) \right) 
\eqend
and has to be supplemented by the constraint (= Gauss' law) 
\eq
\label{Gauss}
G(x)=-\partial_1 E(x)+e\rho(x) \simeq 0 \: . 
\eqend
We note in passing that we also have two Noether currents,  
a vector current $(J_{\iv}^\mu)$ and an axial current 
$(J^\mu_{\ia})$. 
These currents are given by 
\eqa
J_{\iv}^0=\rho,\quad J_{\iv}^1=J\nonu
J^0_{\ia}=J ,\quad J^1_{\ia}= - \rho  
\eqaend
and formally (i.e.\ prior to quantization) 
obey continuity equations, 
$\partial_\mu J_{\iv}^\mu = \partial_\mu J^\mu_{\ia} = 0$. 
We also introduce more general kinds of fermion currents
\eq
\rho_{j,\sigma}(x) = \psi^\dag_{j,\sigma}(x) 
\psi_{j,\sigma}(x) 
\eqend
which all are observables of interest for our model.

\subsection*{4. Construction of the model} 
We now outline a rigorous construction of this model using the
representation theory of loop groups \cite{LS,GLR}. 
This construction amounts to
representing the (Fourier modes) of the field operators
$\psi_{j,\sigma}^{(*)}$, $A_1$ and $E$, and the 
observable algebra of the model
by closed operators on a Hilbert space $\cH$ 
such that the Hamiltonian $H$ is represented by
a self-adjoint operator on $\cH$. In this construction it is crucial to
establish gauge- and Lorentz invariance. As we shall see, this will
lead us to the condition in Eq.\ \Ref{m345}.

The essential physical requirement determining the construction of the
model and implying a non-trivial vacuum structure is positivity of
the Hamiltonian $H$ on the physical Hilbert space.
As is well-known, this  forces one to use a
non-trivial representation of the field operators of the model. The
essential simplification in (1+1) (and not possible in higher)
dimensions is that one can use a quasi-free representation for the
fermion field operators corresponding to ``filling up the Dirac sea''
associated with the {\em free} fermion Hamiltonian $H_0$, and for the
photon operators one can use a naive boson representation. We now
describe this representation in more detail.

In the following it is
convenient to work in Fourier space. We first introduce some 
useful notation which is such that in all equations
the limit $L\to \infty$ is obvious. The Fourier space for even
(periodic) and odd functions is
\nonueqa
\dLam\equiv \left\{\left. p=\f{2\pi}{L} n \right| n\in\Z\right\}
\quad \mbox{ and } \quad 
\dLamF\equiv \left\{\left. q =\f{2\pi}{L} \left(n+\f{1}{2}\right)
\right| n\in\Z\right\} \: , 
\nonueqaend
respectively. For functions $\hat f$ on Fourier space we write
$
\dInt \Del p \hat f(p) \equiv \sum_{p\in\dLam} \f{2\pi}{L} \hat f(p)
$
and similarly for $\dLamF$, and the corresponding 
$\del$-function satisfying $\dInt\Del
p'\, \ddel(p-p')\hat f(p')=\hat f(p)$ is
$\ddel(p-p') \equiv \f{L}{2\pi} \del_{p,p'}$. Our conventions for 
the Fourier transformed operators are, 
\eq
\label{10a}
\hat\psi^{(\dag)}_{j,\sigma}(q) =
\Int \f{\dd{x}}{\sqrt{2\pi}}\, \psi^{(\dag)}_{j,\sigma}(x)
\ee{\stackrel{(+)}{-}\ii qx}
\quad (q\in\dLamF)
\eqend
(i.e.\ we use anti-periodic boundary conditions for the fermions),
\eq
\label{A1}
\hat A_1(p) = \Int \f{\dd{x}}{2\pi}\,
A_1(x) \ee{-\ii px} \quad (p\in\dLam)
\eqend
and in the other cases
\eq
\label{other}
\hat Y(p) = \Int \dd{x}\,
Y(x) \ee{-\ii px}\quad (p\in\dLam)\quad \mbox{ for
$Y=E,\rho,J,\rho_{j,\sigma}$ etc.}
\eqend
Therefore the  non-trivial canonical (anti-) commutator 
relations of the field operators become
$
\ccr{\hat A_1(p)}{\hat E(p')} = \ii\ddel(p+p')$
and 
\eqa
\label{car} 
\car{\hat\psi_{j,\sigma}(q)}{\hat\psi^\dag_{j',\sigma'}(q')}
&=& \del_{\sigma,\sigma'}\del_{j,j'}\ddel(q-q') \nonu
\car{\hat\psi^{(*)}_{j,\sigma}(q)}{\hat\psi^{(*)}_{j',\sigma'}(q')}
&=& 0 \quad \forall p,p',j,j'\: . 
\eqaend

The model will be constructed in the full Hilbert space
$\cH = \cH_{\rm Photon} \otimes \cH_{\rm Fermion}$.

For $\cH_{\rm Photon}$ we take the Fock space
generated by boson field operators $b^{\dag}(p)$ 
and $b^\dag(p)=b(p)^*$, 
$p\in\dLam$, ($*$ is the Hilbert space adjoint)
obeying the commutator relations
\eq
\ccr{b(p)}{b^\dag(p')} = \ddel(p-p'), 
\quad \ccr{b(p)}{b(p')}=0\quad \forall p,p' 
\eqend
and a normalized state $\Omega_{\rm P}\in 
\cH_{\rm Photon}$ such that
\eq
b(p)\Omega_{\rm P} = 0 \quad \forall p \: . 
\eqend
We then set
\eq
\label{photon}
\hat A_1(p) = \f{1}{s}\left( b(p) + b^\dag(-p) \right)
\quad \hat E(p) = -\f{\ii s}{2}\left( b(p)-b^\dag(-p) \right)
\eqend
with a parameter $s$ to be determined later. 
We recall that these requirements 
fix the Hilbert space $\cH_{\rm Photon}$ completely. 
In this setting we now
construct bilinears in the photon field using normal ordering
$\normal{\cdots}$ with respect to the state $\Omega_{\rm P}$, e.g.\ 
$\normal{b(p)b^\dag(p')}=b^\dag(p')b(p)$ for all $p,p'$. 

{}For $\cH_{\rm Fermion}$ we take the Fermion Fock space generated by
operators $\hat\psi_{j,\sigma}(q)$ and  
$\hat\psi^\dag_{j,\sigma}(q)=\hat\psi_{j,\sigma}(q)^*$
obeying the relations Eq.\
\Ref{car} and a normalized 
state $\Omega_{\rm F}\in\cH_{\rm Fermion}$ such that
\eqa
\label{11}
\hat\psi^{}_{+,j}(q) \,\Omega_{\rm F} = 
\hat\psi^{*}_{-,j}(-q) \,\Omega_{\rm F} = 0
\quad \forall q > 0, \forall j \: .
\eqaend
We note that the state $\Omega_{\rm F}$ characterizing this
representation of the fermion field algebra can be
interpreted as Dirac sea associated with the free fermion Hamiltonian
$H_0$. The presence of this Dirac sea requires
normal-ordering $\normal{\cdots}$ of the Fermion bilinears such as
$H_0=\sum_{j} 
\dIntF\Del q \normal{q\,\hat\psi^\dag_j(q)\gamma_3\hat
\psi_j(q)}$ and $\hat \rho_{j,\sigma}(p)$. 
This modifies the naive commutator relations of these operators which
follow from the canonical anti-commutator relations Eq.\ \Ref{car} 
as Schwinger terms show up, see e.g.\ \cite{CR,A,GLR}.
The relevant commutators for us are, 
\eqa
\label{12}
\ccr{\hat\rho_{j,\sigma}(p)}{\hat\rho_{j',\sigma'}(p')} &=&
\sigma\del_{\sigma,\sigma'}\del_{j,j'}p \ddel(p+p')\nonu
\ccr{H_0 }{\hat\rho_{j,\sigma}(p)}&=& 
\sigma p \hat\rho_{j,\sigma}(p) \: .
\eqaend
We also note that
\eq
\label{vacF}
\hat\rho_{j,+}(p)\Omega_{{\rm F}} = \hat\rho_{j,-}(-p)\Omega_{{\rm F}} =
0\quad\forall p>0, \forall j
\eqend
which together with \Ref{12} shows that the $\hat\rho_{j,+}(p)$ (resp.
$\hat\rho_{j,-}(p)$) give a highest (resp.\ lowest) weight
representation of the Heisenberg algebra.

The (Fourier transformed) Gauss' law operators are now well-defined on
$\cH$,  
\eq
\label{14}
\hat G(p) = -\ii p\hat E(p) + e\hat\rho(p)
\eqend
with $\hat\rho(p)=\sum_{j,\sigma}
n_{j,\sigma}\hat\rho_{j,\sigma}(p)$. 
To determine a (possible) gauge anomaly we compute the commutators of
these operators and obtain (we use Eq.\ \Ref{12})
\eq
\ccr{\hat G(p)}{\hat G(p')} = 
e^2 p\ddel(p+p')
\sum_{j=1}^{\Nf}\sum_{\sigma=\pm} \sigma n_{j,\sigma}^2  
 \: .
\eqend
We see that no gauge anomaly occurs if and only if the condition Eq.\
\Ref{m345} holds. In the following we assume that this is the case. 

We now can give a precise meaning to the formal
Hamiltonian $H$ in Eq.\ \Ref{Ham} as follows, 
\eq
\label{Ham1}
H = H_0 + \dInt\Del p \normal{\left(\frac{1}{4\pi}\hat E(p)\hat
E(-p) + e \hat A_1(p) \hat J(-p) + \pi 
\M^2 \hat A_1(p)\hat A_1(-p) \right) } \: .  
\eqend
Similarly as in the Schwinger model we have added a photon mass term 
(= last term on the r.h.s. of Eq.\ \Ref{Ham1}) to restore gauge invariance 
which otherwise would be spoiled by
Schwinger terms. Note that our normalization is such that
this term formally equals $\half\M^2\Int \dd{x} A_1(x)^2$, 
i.e.\ $\M$ can be interpreted as 
photon mass. To determine the parameter $\M^2$ we
compute $\ccr{\hat G(p)}{H}$ and demand this to be always zero. 
By straightforward computation (we use Eq.\ \Ref{12}), 
\eq
\label{mass}
\M^2 = \frac{e^2}{2\pi}
\sum_{j=1}^{\Nf}\sum_{\sigma=\pm} n_{j,\sigma}^2 \: . 
\eqend
We finally have to fix the parameter $s$ in Eq.\ \Ref{photon}. 
We observe that $H$ in Eq.\ \Ref{Ham1} has the form $H=H_{0,{\rm
F}}+ H_{0,{\rm P}}+H_{\rm int}$ where $H_{0,{\rm F}}=H_0$ and
\eq
H_{0,{\rm P}}= \dInt\Del p 
\normal{\left(\frac{1}{4\pi}\hat E(p)\hat
E(-p) + \pi \M^2 \hat A_1(p)\hat A_1(-p) \right) } 
\eqend
are the free fermion- and Photon Hamiltonians. We fix $s$ by requiring
that $H_{0,{\rm P}}$ has the form $\dInt\Del p \, \omega(p) b^\dag(p)b(p)$. 
We thus obtain $s=2\sqrt{\pi|\M|}$ and $\omega(p)= |\M|$.
Then $H_{0,{\rm F}}+ H_{0,{\rm P}}$ obviously is defined 
as self-adjoint, positive operators on $\cH$. 
Our results in this paper imply that also the interacting Hamiltonian 
$H$ is well-defined on $\cH$: the operator $H$ is self-adjoint and bounded 
from below on $\cH$. 

\subsection*{5. Bosonization and gauge fixing I} 
As in the standard Schwinger model we can rewrite the Hamiltonians 
of our models as boson Hamiltonians by using the so-called Kronig 
identity 
$
\hat H_0 = \f{1}{2}\dInt\Del p \,
\normal{
\sum_{j,\sigma}
\hat\rho_{j,\sigma}(p)\hat\rho_{j,\sigma}(-p) }
$
where $\normal{\cdots}$ denotes normal ordering of the
fermion currents (see \cite{GLR} for more details). 
We obtain 
\eq
\label{20}
H = \dInt\Del q\, 
\normal{\left(
 \f{1}{4\pi}\hat E(p) \hat E(-p) + 
\f{1}{2}
\sum_{j=1}^{\Nf}\sum_{\sigma=\pm}
\tilde\rho_{j,\sigma}(p)\tilde\rho_{j,\sigma}(-p)\right)}
\eqend
where we have defined
\eq
\tilde\rho_{j,\sigma}(p)\,  = \hat \rho_{j,\sigma}(p)  
+ \sigma e n_{j,\sigma}\hat A_1(p) \: . 
\eqend
Note that $\ccr{\hat G(p)}{\tilde\rho_{j,\sigma}(p')}=0$ i.e.\ the 
operators $\tilde\rho_{j,\sigma}(p)$ all are gauge invariant
and thus observables for our model.

The Hilbert space $\cH$ which we have obtained still contains
gauge equivalent states i.e.\ states related to each other 
by static gauge transformations. To obtain the 
physical Hilbert space 
$\cH_{\rm phys}$ for our models we have to go through a
procedure called {\em fixing the gauge}
in the physics literature. We will do it in two steps.  
The first step will be done in this paragraph and eliminates 
the `small' gauge 
transformations (i.e.\ the $\ee{\ii\chi}\in\cG$ with $\chi$ smooth and 
periodic functions on space $S^1$). This step is easy since the 
resulting Hilbert space $\cH'_{\rm phys}$ can be identified 
as the sub-Hilbert space of $\cH$ which is annihilated by all the 
Gauss law operators. In a second step, which will be described 
in Paragraph 7 further below, we will account for 
large gauge transformations, too.  

For the first step we recall that 
the only gauge invariant degree of freedom of the Photon field at fixed
time is the holonomy $\Int \dd{x}\, A_1(x)$. 
Due to the absence of a gauge anomaly we can therefore 
impose the gauge condition
\eq
\label{gf1}
\hat A_1(p) = \del_{p,0} Y 
\eqend
and solve the Gauss' law $\hat G(p)\simeq 0$ (cf.\ Eq.\ \Ref{14}) as
\eq
\label{gf2}
\hat E(p) = 
\f{e\hat\rho(p)}{\ii p} \quad {\mbox{for $p\neq 0$}}.
\eqend
This determines all components of $\hat E$ except the one 
conjugate to $Y$: 
\eq
\label{gf3}
\hat E(0) = \f{L}{2\pi\ii } \f{\partial}{\partial Y} \: . 
\eqend
After
that we are left with the ($p=0$)-component of Gauss' law, {\em viz.}
\eq
eQ_{\iv}\simeq 0,\quad Q_{\iv} = \hat\rho(0) = 
\sum_{j=1}^{\Nf}\sum_{\sigma=\pm} n_{j,\sigma} 
\hat\rho_{j,\sigma}(0) \: . 
\eqend
We thus can identify the  Hilbert space of states invariant under
all small gauge transformations as   
$\cH'_{\rm phys}= L^2(\R,{\rm d}Y)\otimes {\cal H}'_{\rm 
Fermion}$ where ${\cal H}'_{\rm Fermion}$ is the zero charge 
sector of the fermionic Fock space. Moreover, by inserting 
Eqs.\ \Ref{gf1}--\Ref{gf3}, each gauge invariant operator on 
$\cH$ becomes an operator on $\cH'_{\rm phys}$. Especially 
the Hamiltonian becomes $H= \frac{2\pi}{L} 
\sum_{p\geq 0} \normal{h_p}$ where 
\eq
h_0 =  
-\f{1}{4\pi} \left(\f{L}{2\pi}\right)^2 
\f{\partial^2}{\partial Y^2} +
\f{1}{2}
\sum_{j=1}^{\Nf}\sum_{\sigma=\pm}
\left(\hat\rho_{j,\sigma}(0)+\sigma n_{j,\sigma}eY
\right)^2 
\eqend
and 
\eqa
\label{24}
\normal{h_p} \: = \, 
\normal{\left(  \f{e^2}{2\pi p^2} \hat\rho(-p)\hat\rho(p)
 + \sum_{j=1}^{\Nf}\sum_{\sigma=\pm}
\hat\rho_{j,\sigma}(-p)\hat\rho_{j,\sigma}(p) \right)}  \: . 
\eqaend
This completes the first step of the gauge fixing procedure. 
As mentioned we have not yet accounted for the existence of the
large gauge transformations. We will come back to this in 
Paragraph 7 below.

\subsection*{6. Solution of the model}
The operators $\normal{h_p}$ obviously all commute with each other. 
We thus can diagonalize the Hamiltonian $H$ by 
diagonalizing each term $\normal{h_p}$ separately. 

We first consider $\normal{h_p}$ with $p>0$. We introduce the boson operators 
\eq
\label{crho}
c_j(p) = \left\{\bma{cc} \f{1}{\sqrt{|p|}}\hat\rho_{j,+}(p) & \mbox{ for
$p>0$}\\
\f{1}{\sqrt{|p|}}\hat\rho_{j,-}(p) & \mbox{ for $p<0$} \ema\right.
\eqend
and $c^\dag_{j}(p) = c_{j}(p)^*=c_{j}(-p)$, so that
$\ccr{c_j(p)}{c^\dag_{j'}(p')}=\frac{L}{2\pi}\del_{j,j'}\ddel(p-p')$
and $c_j(p)\Omega=0$ for all $j,j'$ and $p,p'\neq 0$. 
We also find it convenient to define
\eq
c_j(p)=c_j(p),\quad c_{\Nf+j}(p)=c^\dag_j(p)\quad j=1,\ldots \Nf  \: ,  
\eqend
to fix $p>0$, and suppress the argument $p$ in the following. 
Then $\ccr{c_{j}}{c^\dag_{j'}}=\frac{L}{2\pi}\del_{j,j'}q_{j}$ 
where $q_{j}=1$ and $q_{\Nf+j}=-1$ for $j=1,\ldots,\Nf$.

We then can write $\normal{h_p}$ in
matrix notation as $\normal{\vc^\dag \cdot \mOm  \vc}$ 
where we defined $\vc^\dag \, = (c_1^\dag,\ldots, c_{2\Nf}^\dag)$ 
etc.\ and $\mOm$ is the $2\Nf\times 2\Nf$ matrix 
\eq
\mOm = p\id + \f{e^2}{2\pi p} \vn \otimes\vnt 
\eqend
where $\vnt=(n_1,\ldots,n_{2\Nf})$ with $n_j=n_{j,+}$ 
and $n_{\Nf+j}=n_{j,-}$ for $j=1,\ldots,\Nf$ 
($\id$ is the $2\Nf\times 2\Nf$ unit matrix
here, and we use a standard tensor notation; the $^T$ denotes the 
matrix transpose). Defining $h_p =\vc^\dag \cdot \mOm  \vc$
we get $\normal{h_p}\,  = h_p -<\Omega,h_p\Omega>$ with 
\eq
<\Omega,h_p\Omega> = \f{L}{2\pi}\sum_{j=1}^{\Nf} \left(p +  
\f{e^2}{2\pi p} n_{j,-}^2
\right)  =
 \f{L}{2\pi}\left( \Nf|p| +  \f{M^2}{2 p}\right)  
\eqend
(we used Eqs.\ \Ref{mass} and \Ref{m345}). 

We now can diagonalize $h_p$ by finding a boson Bogoliubov transformation 
$\vC =\mU\vc$
($\mU$ some $2\Nf \times 2\Nf$ matrix) so that the operators $C_j$ obey 
the same relations as the $c_j$ and are such 
that $h_p = \vC^\dag \cdot \mom   \vC = 
\sum_{j=1}^{2\Nf} \omega_j C_j^\dag C_j$. It is easy to see that 
these two conditions are equivalent to
\eq
\label{dg}
\mU \mq \mU^*=\mq
\quad\mbox{ and } \quad 
(\mU^{-1})^* \mOm \mU^{-1}=\mom 
\eqend
where $\mq=diag(q_1,\ldots,q_{2\Nf})$ and $\mom=diag(\omega_1,\ldots, 
\omega_{2\Nf})$ ($^*$ and $^{-1}$  is the matrix
adjungation and matrix inverse, respectively). To  
solve this somewhat unconventional diagonalization problem we note
that Eq.\ \Ref{dg} implies
\eq
\label{sdg}
\mom^2 = 
(\mq\mom)^2  = 
\mU (\mq\mOm)^2  \mU^{-1} \: ,
\eqend
and $\mU$ is determined from this equation up to transformations
 $\mU\to \mV\mU$
where $\mV$ commutes with $\mom^2$. We shall see that Eq.\ \Ref{sdg}
corresponds to a standard diagonalization problem: the matrix
$(\mq\mOm)^2$ is self-adjoint and thus can be
diagonalized by a {\em unitary} matrix $\mUt$. We
thus can solve the problem in Eq.\ \Ref{dg} by 
first determining $\mUt$ and then making the ansatz
$\mU=\mV\mUt$ with $\mV$ commuting with $\mom^2$. From 
Eq.\ \Ref{dg} we then get the following condition, 
\eq
\label{sdg1}
\mV\mUt \mq \mOm  \mUt^* \mV^{-1}=
\mq\mom 
\eqend
which again is a standard diagonalization problem and 
will allow us to determine $\mV$. 

We now compute $\mom^2$ using Eq.\ \Ref{sdg}. We write 
$\mq \mOm = p\mq + \frac{\M^2}{p}\ve_{\Nf+1}\otimes\vet_1$ 
where we define $\ve_1\, =\vn/|\vn|$ ($|\vn|=\sqrt{\vnt\cdot\vn}$) 
and $\ve_{\Nf+1} \, : 
=\mq\ve_1$ (we used $|\vn|=|\mq\vn|$ 
and $\frac{e^2}{2\pi}|\vn|^2 = \M^2$). 
It is now crucial to observe that 
the condition Eq.\ \Ref{m345} is equivalent to 
$
\vn^T\cdot\mq \vn=0 
$ i.e.\ {\em if there is no gauge anomaly the two vectors $\ve_{1}$ and 
$\ve_{\Nf+1}$ are orthonormal}. Moreover, in this case we  
can extend these vectors to a complete
orthonormal real basis $\{\ve_j\}_{j=1}^{2\Nf}$ in $\R^{2\Nf}$ so that 
$\mq\ve_j=\ve_{\Nf +j}$ for $j= 1,\ldots, \Nf$. We thus obtain
$$
(\mq\mOm)^2   = p^2 \id + 
\M^2 \left(\ve_1\otimes\vet_1 + \ve_{\Nf+1}\otimes\vet_{\Nf+1}\right) 
$$
from which we can immediately read off $\mUt$ and the matrix elements
of $\mom^2$: denoting as $\vE_j$ the standard basis in $\R^{2\Nf}$
(i.e.\ $(\vE_j)_{j'}=\del_{j,j'}$) we have 
\eq
\label{Ut}
\mUt = \sum_{j=1}^{2\Nf} \vE_j \otimes\vet_j \: , 
\eqend
and $\omega_j^2= p^2+\M^2 $ and for $j=1,\Nf+1$ and $\omega_j^2= p^2$ 
otherwise. We then compute
\eq
\label{hm}
\mUt  \mq\mOm\mUt^* = 
\sum_{j=1}^{\Nf}
\left(\f{\M_j^2}{p}\vE_{\Nf+j}\otimes\vEt_j +
p \vE_{\Nf+j}\otimes\vEt_j +  p \vE_j\otimes\vEt_{\Nf+j}\right) 
\eqend  
with $\M_1=\M$ and $M_{j\neq 1} =0$, which shows that we can determine  
$\mV$ by diagonalizing $2\times 2$ matrices of the form 
$\left(\bma{cc}  0 & p \\ p+ \f{\M_j^2}{p} & 0 \ema\right)$. We find
\eq
\label{mV}
\mV = \sum_{j=1}^{2\Nf} \vE_j \otimes \vFt_j 
\eqend
where
\eq
\label{vFj}
\vF_{j,\Nf+j}=
\frac{p\vE_j\pm \sqrt{p^2+\M_j^2} \vE_{\Nf+j} }{
\sqrt{2p^2 + \M_j^2}}  \: . 
\eqend
Thus
\eq
\label{omj}
\omega_j=-\omega_{\Nf+j}=\sqrt{p^2+\M_j^2} \quad 
\mbox{ for $j=1,\ldots,\Nf$} \: . 
\eqend
We thus obtain 
\eq
\label{hpd}
h_p = \sum_{j=1}^{\Nf} \sqrt{p^2+\M_j^2}
\left( C_j(p)^\dag C_j(p) + C_{j}(-p)C_{j}(-p)^\dag
\right)  
\eqend
where we used $C_{\Nf +j}(p)=C_j^\dag(-p)$. This completes the 
diagonalization of the operators $\normal{h_p}$.

We are left to diagonalize $h_0$. We note that this is the 
part of the Hamiltonian which contains the operators $Y$ and 
$\hat \rho_{j,\sigma}(0)$ which are not 
invariant under the large gauge
transformation $\ee{\ii x /L}$ but transform as follows, 
\eqa
\label{lgt}
Y &\to& Y-\frac{1}{e} \nonu
\hat \rho_{j,\sigma}(0) &\to & \hat \rho_{j,\sigma}(0) +
\sigma n_{j,\sigma} \: . 
\eqaend
It is therefore not immediately obvious that $h_0$ is indeed 
invariant under the large gauge transformations. 
To make this invariance manifest 
we note that the operator
$Q_{\ia}=\sum_{j,\sigma}\sigma n_{j,\sigma}\hat\rho_{j,\sigma}(0)$ 
changes under the transformations Eq.\ \Ref{lgt} as $Q_{\ia}\to Q_{\ia} + 
|\vn|^2$,  and therefore  
\eq
\tilde Y \:  : = Y + \frac{1}{e|\vn|^2}Q_{\ia}
\eqend
is indeed invariant. By straightforward computations we obtain 
\eq
h_0 = \frac{|\M|}{2}\left( C_1(0)^\dag C_1(0) + C_1(0)C_1(0)^\dag 
\right) + \cC
\eqend
where $C_1^{(\dag)}(0) = \stackrel{(-)}{+}
\frac{L}{4\pi} (\pi|\M|)^{-1/2}\frac{\partial}{\partial \tilde Y}
+ (\pi|\M|) ^{1/2} \, \tilde Y $
and 
\eq
\label{CC}
\cC = 
\frac{1}{2}\left( 
\sum_{j=1}^{\Nf}\sum_{\sigma=\pm}
\hat\rho_{j,\sigma}(0)^2 - \frac{1}{|\vn|^2}Q_{\ia}^2  
\right) 
\eqend
all are invariant under the transformations Eq.\ \Ref{lgt}.
We also have $\ccr{C_1(0)}{C_1^\dag(0)}=\frac{L}{2\pi}$ and 
thus see that $h_0$ is essentially a harmonic oscillator 
Hamiltonian.

We can combine our results above in a compact form as follows,  
\eq
\label{nice}
H= \sum_{j=1}^{\Nf} \dInt \Del p \sqrt{p^2+\M_j^2}\, 
C_j^\dag(p)C_j(p) +\frac{2\pi}{L}\cC + \cE_0 \: ,
\eqend
with the constant 
\eq
\label{cE}
\cE_0 =
\sum_{j=1}^{\Np}\left( 
\frac{|M_j|}{2} + 
\frac{L}{2\pi} \hat{\int}_{p>0} \Del p\, \left( \sqrt{p^2+\M_j^2} - 
(p+\frac{\M_j^2}{2p})  \right) \right)
\eqend
(note that actually only the term with $j=1$ contributes to $\cE_0$). 
This shows explicitely that the Hamiltonian $H$ is self-adjoint and 
bounded from below, and since $\cC$ is non-negative (we will show 
this further below) the constant $\cE_0$ is equal to the the ground 
state energy. Moreover, we also see explicitly that our model 
has a relativistic spectrum: the physical degrees of freedom 
correspond to one massive- and $\Nf-1$ massless boson fields. 

We now briefly describe how to construct a groundstate 
for our model. We note that with the explicit formulas 
given above it is straightforward to construct a unitary 
operator $\cU(p)$ in terms of the operators 
$c^{(\dag)}_j(p)$ such that 
$C_j(p)=\cU(p)c_j(p)\cU(p)^*$. 
Moreover, one can check that $\cU =\prod_{p>0}\cU(p)$  
defines a unitary operator on $\cH'_{\rm Fermion}$. 
It is then easy to see that the state
$ \phi_0(Y) \cU \Omega_{\rm F}\in\cH'_{\rm phys}$
with $\phi_0(Y)\propto 
\exp(-\pi |\M| \frac{2\pi}{L}Y^2)$  
is annihilated by all operators $C_j(p)$ and $\cC$ 
and thus a ground state 
of $H$. We thus have found {\em one} groundstate for our model. 
However, this state is 
highly degenerate, and it is actually not invariant under 
the large gauge transformations Eq.\ \Ref{lgt}. 
In the next paragraph we will discuss this issue 
in more detail. 

We finally mention that in a complete solution 
of the model one also needs to find the Green's functions 
i.e.\ vacuum expectation values of gauge invariant combinations 
of the fermion- and photon field operators. All these Green's 
functions can be computed explicitly by using 
the so-called boson-fermion correspondence 
which allows to write the fermion field operators in terms of 
the fermion currents. These computations are  similar to 
the ones for the usual Schwinger model (see e.g.\ \cite{GLR}) 
but beyond the scope of the present paper.

\subsection*{7. Gauge fixing II: Vacuum structure and all that} 
We now describe the structure of the Hilbert space of our model
and then perform the second step of the gauge fixing procedure 
described already above. 

We first recall the well-known 
structure of the fermion Fock space $\cH_{\rm Fermion}$: 
this space can be generated by the
fermion current $\hat\rho_{j,\sigma}(p)$, $p\neq 0$, 
together with unitary operators $R_{j,\sigma}$ which obey the 
relations 
\eq
R_{j,\sigma}^{-1} \hat\rho_{j',\sigma'}(p)
R_{j,\sigma} = \hat\rho_{j',\sigma'}(p) + 
\del_{p,0}\del_{j,j'}\sigma\del_{\sigma,\sigma'} 
\eqend 
and which interpolate between different sectors labeled
by the eigenvalues of the charge operators. 
Thus for all 
$\vm=(m_{1,+},\ldots,m_{\Nf,+}, m_{1,-},\ldots, m_{\Nf,-})\in\Z^{2\Nf}$,  
the operators
\eq
\RR^{\vm} = R_{1,+}^{m_{1,+} } \cdots R_{\Nf,+}^{m_{\Nf,+} }
R_{1,-}^{m_{1,-} } \cdots 
 R_{\Nf,-}^{m_{\Nf,-} }
\eqend
commute with all $h_{p>0} $ and $\cU$, and if $\Psi$ is a vector 
in $\cH_{\rm Fermion}$ with $\hat\rho_{j,\sigma}(0)\Psi=0$  
$\forall j,\sigma$ then
$\hat\rho_{j,\sigma}(0)\RR^{\vm}\Psi=m_{j,\sigma}\RR^{\vm}\Psi$, which implies
that $\RR^{\vm}\Psi$ is an eigenvector of $Q_{\iv}$ and 
$Q_{\ia}$ with eigenvalues 
$\vnt\cdot \vm$ and $\vnt\cdot\mq\vm$, respectively. This implies that  
if such a vector $\Psi$ is also an eigenstate of all $h_{p>0}$ and 
$\phi\in L^2(\R)$ an eigenstate of $h_0$ then all states
$$
\phi(\tilde Y)\RR^{\vm}\Psi = 
\phi\left(Y+\frac{\vn\cdot \mq\vm}{e|\vn|^2} \right)\RR^{\vm}\Psi
$$ 
with $\vnt\cdot\vm=0$ (= charge zero condition) are eigenstates of $H$ 
with eigenvalues which are of the form $ \cE + c(\vm) $ and differ 
only by the contribution from $\cC$ Eq.\ \Ref{CC}, $c(\vm)= 
\frac{2\pi}{L |\vn|^2}(|\vm|^2|\vn|^2 - (\vnt\cdot\mq\vm)^2 )$.
Note that the latter is always non-negative due to the Cauchy-Schwartz 
inequality. We thus see that all these states with 
$\vm= k\mq\vn$ ($k$ integer) are degenerate, and especially all states 
$\phi_0\left(Y + \frac{k}{e} \right)\RR^{k\mq\vn}\cU\Omega$
are groundstates for our models. This degeneracy actually 
is explained by the invariance under large gauge transformation 
Eq.\ \Ref{lgt}: This transformation acts on states in 
$\cH'_{\rm phys}$ as
$
\phi(Y) \Psi\to \phi\left(Y+\frac{1}{e} \right) \RR^{\mq\vn}\Psi \: . 
$

We now come to the second step of our gauge fixing procedure. 
Our discussion above implies that the states 
\eq
[\Psi,\phi]^\theta(Y) = \frac{1}{\sqrt{2\pi}}
\sum_{k\in\Z} \ee{\ii k \theta } 
\phi\left(Y+\f{k}{e}\right) \RR^{k\mq\vn}\Psi   
\eqend 
($\theta$ real) 
have simple transformation properties
under large gauge transformation Eq.\ \Ref{lgt}, 
$[\Psi,\phi]^\theta\to \ee{\ii\theta }[\Psi,\phi]^\theta$. 
We thus can define $\cH_{\rm phys}$ as the vector space spanned by all 
$[\Psi,\phi]^0(Y)$. However, this does not yet make 
$\cH_{\rm phys}$ into a Hilbert space: a simple computation shows that
$$<[\Psi_1,\phi_1]^\theta,[\Psi_2,\phi_2]^{\theta'}>_{\cH_{\rm phys} } 
= 
\del(\theta-\theta')<[\Psi_1,\phi_1],[\Psi_2,\phi_2]>_{\rm phys}$$ 
where 
$
<[\Psi_1,\phi_1],[\Psi_2,\phi_2]>_{\rm phys}  = 
<\Psi_1,\Psi_2>_{\cH_{\rm Fermion}}<\phi_1,\phi_2>_{L^2(\R)} 
$  
is independent of $\theta$ and $\theta'$. We see that 
the vectors $[\Psi,\phi]^\theta(Y)$ are actually not contained in
$\cH'_{\rm phys}$ (they do not have finite norm). 
The remedy of this problem is a simple 
multiplicative regularization i.e.\ `dropping 
the infinite constant $\del(0)$'. This is equivalent 
to using $<\cdot,\cdot>_{\rm phys}$ as inner product in 
$\cH_{\rm phys}$ which is well-defined. 
This completes the construction of the model.

\subsection*{8. Generalization to an arbitrary number of photons} 
We now describe how the above results 
generalize to a large class of models with $\Np$ 
different photon fields $A_\mu^\alpha$ where $\alpha=1,\ldots,\Np$. 
These models are (formally) given by                          , 
\eq
\cL = -\frac{1}{4} \sum_{\alpha=1}^{\Np} 
F^\alpha_{\mu\nu}(F^\alpha)^{\mu\nu} + 
\sum_{j=1}^{\Nf}\bar\psi_j \gamma^\mu[-\ii\partial_\mu + 
\sum_{\alpha=1}^{\Np}
e^\alpha (n^\alpha_{j,+}P_+ +n^\alpha_{j,-}P_-) A^\alpha_\mu ] 
\psi_j 
\eqend
where $F^\alpha_{\mu\nu}= 
\partial_\mu A^\alpha_\nu-\partial_\nu A^\alpha_\mu$ 
and the charge units $e^\alpha$ corresponding to the 
different gauge fields can be different. This model 
obviously is invariant under transformations  belonging to 
the gauge group $\cG=C^\infty(\R\times S^1; \U(1)^{\Np} )$, 
$
A^\alpha_\mu \to A^\alpha_\mu - 
\frac{1}{e^\alpha}\partial_\mu\chi^\alpha
$
etc., and as before the existence of large gauge transformations 
requires all the $n^\alpha_{j,\sigma}$ to be integers. 
The canonical procedure and the construction of the models with 
$\Np=1$ generalize with minor changes: now we have 
$\Np$ copies of the photon fields and correspondingly $\Np$ copies of the 
Gauss' law operators etc. To be more specific: We now have
$\ccr{\hat A^\alpha_1(p)}{\hat E^\beta (p')} = \ii\del^{\alpha,\beta}
\ddel(p+p')$, a Hamiltonian
\eq
H = H_0 + \dInt\Del p \sum_{\alpha=1}^{\Np}\normal{ \left( 
 \frac{1}{4\pi}\hat E^\alpha(p)\hat E^\alpha(-p) + 
e^\alpha \hat A^\alpha_1(p) \hat J^\alpha(-p)   + 
\pi \hat A^\alpha_1(p) \sum_{\beta=1}^{\Np} \Upsilon^{\alpha\beta} 
\hat A^\beta_1(-p) 
\right)}\: , 
\eqend
and  Gauss' law operators
$\hat G^\alpha(p) = -ip\hat E^\alpha(p) + \hat\rho^\alpha(p)$
where
\eq
\hat\rho^\alpha(p) = \sum_{j,\sigma}n_{j,\sigma}^\alpha
\hat\rho_{j,\sigma}(p)\,   ,\qquad
\hat J^\alpha(p) = \sum_{j,\sigma} \sigma n_{j,\sigma}^\alpha
\hat\rho_{j,\sigma}(p) 
\eqend
with $\hat\rho_{j,\sigma}(p)$ as before. 
The model has no gauge anomalies if and only if all
commutators $\ccr{\hat G^\alpha(p)}{\hat G^\beta(p')}$ vanish, 
and similarly as for $\Np=1$ we obtain the conditions in Eq.\ 
\Ref{m345_1}. Note that the mass term we have to add to
the naive Hamiltonian depends on a $\Np\times \Np$-matrix 
$\Upsilon^{\alpha\beta}$ which is determined such that $H$ commutes 
with all $\hat G^\alpha(p)$. We thus obtain
\eq
\label{mcM}
\Upsilon^{\alpha\beta} = 
 \frac{e^\alpha e^\beta}{2\pi}
\sum_{j=1}^{\Nf}\sum_{\sigma=\pm} n_{j,\sigma}^\alpha 
n_{j,\sigma}^\beta 
\: . 
\eqend
We note already here that 
${\bf\Upsilon}=(\Upsilon^{\alpha\beta})_{\alpha,\beta=1}^{\Np}$ 
is a self-adjoint, real, non-negative $\Np\times\Np$ matrix,
and therefore we can write 
\eq
\label{mcM1}
\Upsilon^{\alpha\beta} = \sum_{\gamma=1}^{\Np}
\M_\gamma^2 (a_\gamma)^\alpha(a_\gamma)^\beta
\eqend
where $(a_\gamma)^\alpha$ are the components of the 
orthonormal eigenvectors of 
${\bf\Upsilon}$ and $\M_\gamma^2$ the corresponding eigenvalues. 
For later convenience we assume that 
$\rk({\bf\Upsilon})=\Np$ i.e.\ that all the $\M_\gamma^2$
are non-zero.\footnote{We believe that this assumption
could be easily dropped.}

The generalization of the 
representation Eq.\ \Ref{photon} of the photon fields etc.\
generalizes in a straightforward manner (we only mention
that the generalization of the free Photon Hamiltonian now becomes 
$H_{0,{\rm P}}= \sum_{\alpha}\dInt \Del p \, |\M_\alpha| 
b_\alpha^\dag(p) b_\alpha(p)$ where $\M_\alpha^2$ are the 
eigenvalues of the matrix ${\bf\Upsilon}$). 
Moreover, also the generalization of 
Eq.\ \Ref{20} is obvious where the gauge invariant 
currents now are 
$
\tilde\rho_{j,\sigma}(p)\,  = \hat \rho_{j,\sigma}(p)  
+ \sigma \sum_{\alpha=1}^{\Np} 
e^\alpha n^\alpha_{j,\sigma}\hat A^\alpha_1(p) 
$ (note that due to Eq.\ \Ref{m345_1} these latter 
currents indeed commute with all Gauss law operators). 

We now come to the solution of the models: the gauge 
fixing condition generalizing the one in Eq.\ 
\Ref{gf1} obviously is
$
\hat A^\alpha_1(p) = \del_{p,0} Y^\alpha 
$. Imposing that condition, the Hilbert space of states invariant under
all small gauge transformations becomes    
$\cH'_{\rm phys}= L^2(\R^{\Np},{\rm d}Y^1\cdots 
{\rm d}Y^{\Np} )\otimes {\cal H}'_{\rm  Fermion}$
where the zero charge sector in the
fermion Fock space now is defined such that 
\eq
Q^\alpha_{\iv} = 
\sum_{j=1}^{\Nf}\sum_{\sigma=\pm} n^\alpha_{j,\sigma} 
\hat\rho_{j,\sigma}(0) = 0 \quad \forall \alpha=1,
\ldots,\Np 
\eqend
on  ${\cal H}'_{\rm  Fermion}$ (again, these latter
conditions come from the  ($p=0$)-components of 
Gauss' law). Moreover, the Hamiltonian can again 
be written as $H= \frac{2\pi}{L} \sum_{p\geq 0} \normal{h_p}$. 

For $p>0$ we can use the  
matrix notation introduced in Paragraph 6 and write 
$h_p= \vc^\dag \cdot \mOm  \vc$ where 
$
\mOm = p\id + \frac{1}{p}\mM$ with
\eq
\label{mM}
\mM = \sum_{\alpha=1}^{\Np}
\f{e^\alpha e^\alpha}{2\pi } \vn^\alpha \otimes (\vn^\alpha)^T 
\eqend
and $(\vn^\alpha)^T=(n^\alpha_{1,+},\ldots,n^\alpha_{\Nf,+},
n^\alpha_{1,-},\ldots,n^\alpha_{\Nf,-})$ etc.\ as before. 
We now show how to diagonalize these Hamiltonians, following 
the method explained in Paragraph 6: we compute
$
(\mq\mOm)^2   = p^2 \id + \mM + \mq \mM \mq
$
where we used $\mM\mq\mM=0$ which follows from 
Eq.\ \Ref{m345_1}. We now observe that $\mM$ is a self-adjoint, real,  
non-negative matrix, and it therefore 
can be written as
\eq
\mM = \sum_{j=1}^{\Nf} \M_j^2 \ve_j \otimes \vet_j  
\eqend
where the $\ve_j$ are orthonormal and 
$\M_1^2\geq \M_2^2\geq \ldots \geq \M_{\Nf}^2$. 
The $\M_j^2$ and $\ve_j$ can computed by diagonalizing
the $\Nf\times \Nf$ matrix $\mM $. We also observe that 
the {\em non-zero} eigenvalues $\M_j^2$  of the matrix $\mM $
are identical with the non-zero eigenvalues of the 
matrix ${\bf\Upsilon}=(\Upsilon^{\alpha\beta})_{\alpha,\beta=1}^{\Np}$ 
defined in Eq.\ \Ref{mcM}. Thus only the $\M_j^2$ with 
$j=1,\ldots,\rk({\bf\Upsilon})=\Np$ are non-zero. 
(To see this, note that if $(a_j)^\alpha$ are the components
of an eigenvector of  ${\bf\Upsilon}$ with non-zero eigenvalue 
$\M^2_j$, then 
$\vf_j = \sum_{\alpha}e^\alpha (a_j)^\alpha \vn^\alpha$
is an eigenvector of the matrix $\mM$ with the same 
eigenvalue $\M^2_j$. One can also check easily that
the $\vf_j$ span a space of dimension equal to the rank
of the matrix ${\bf\Upsilon}$ which we assumed to be 
equal to $\Np$.) 

Defining 
$\ve_{\Nf+j} =\mq \ve_{j}$ we obtain 
a complete orthonormal basis
in $\R^{2\Nf}$ (orthogonality of the $\ve_{j\leq\Nf}$ and 
$\ve_{j>\Nf}$ again follows from Eq.\ \Ref{m345_1}), and 
we can write
$$
(\mq\mOm)^2   = p^2 \id + \sum_{j=1}^{\Np}\M_j^2 
\left(\ve_j\otimes\vet_j + \ve_{\Nf+j}
\otimes\vet_{\Nf+j}\right) \: . 
$$
Thus $\mUt$ Eq.\ \Ref{Ut} diagonalizes the matrix $(\mq\mOm)^2$, 
and the eigenvalues of this latter matrix are 
$\om_j^2 = p^2 + \M_j^2$ where $\M_{\Nf+j}^2 = \M_j^2$. 
It is then easy to check that all the Eqs.\ \Ref{hm}--\Ref{omj}
remain true also in the present case, and 
one finally obtains a representation of $h_p$ as in 
Eq.\ \Ref{hpd}.

We now turn to $h_0$. After some computations we obtain, 
\eq
h_0 =  
-\f{1}{4\pi} \left(\f{L}{2\pi}\right)^2 
\sum_{\alpha=1}^{\Np} \f{\partial^2}{\partial (\tilde Y^\alpha) ^2} +
\f{1}{2} \sum_{\alpha,\beta=1}^{\Np} \Upsilon^{\alpha\beta} \tilde Y^\alpha 
\tilde Y^\beta + \cC
\eqend
where $\tilde Y^\alpha = Y^\alpha + \sum_\beta
(\Upsilon^{-1})^{\alpha\beta} Q_{\ia}^\beta$, $Q_{\ia} = \sum_{j,\sigma} 
\sigma n^\alpha_{j,\sigma}\hat\rho_{j,\sigma}(0)$, 
and 
\eq
\label{CC_1}
\cC = 
\frac{1}{2}\left( 
\sum_{j=1}^{\Nf}\sum_{\sigma=\pm}\hat\rho_{j,\sigma}(0)^2 - 
\sum_{\alpha,\beta=1}^{\Np} 
Q^\alpha_{\ia} (\Upsilon^{-1})^{\alpha\beta} 
Q^\beta_{\ia}  
\right) \: . 
\eqend
Note that $\tilde Y^\alpha$, $\cC$ and $h_0$ all are
invariant under the large gauge transformations 
\eqa
\label{lgt_1}
Y^\alpha &\to& Y^\alpha-\frac{k^\alpha}{e^\alpha} \nonu
\hat \rho_{j,\sigma}(0) &\to & \hat \rho_{j,\sigma}(0) +
\sum_{\alpha=1}^{\Np} \sigma n^\alpha_{j,\sigma} k^\alpha
\eqaend
where $\vk=(k^1,\ldots,k^{\Np})\in\Z^{\Np}$. 

Introducing $Z_j = \sum_{\alpha}(a_j)^\alpha 
\tilde Y^\alpha$ where $(a_j)^\alpha$ the components
of the eigenvectors of the matrix ${\bf\Upsilon}$ (cf.\
Eq.\ \Ref{mcM1}) and 
$C_j^{(\dag)}(0) = \stackrel{(-)}{+}
\frac{L}{4\pi} (\pi|\M_j|)^{-1/2}\frac{\partial}{\partial Z_j}
+ (\pi|\M_j|) ^{1/2} \, Z_j $ we can write
\eq
h_0 = \sum_{j=1}^{\Np}
\frac{|\M_j|}{2}\left( C_j(0)^\dag C_j(0) + C_j(0)C_j(0)^\dag 
\right) + \cC \: . 
\eqend
Again we can combine our results and write $H$ as in 
Eqs.\ \Ref{nice}--\Ref{cE}, and we see explicitly that
our model has a relativistic spectrum: we have $\Np$ massive 
and $(\Nf-\Np)$ massless bosons.

It is straightforward to extend the construction of the unitary operator
$\cU$ on  $\cH_{\rm Fermion}'$ diagonalizing all the $h_p$ and then 
check that $ \phi_0(Y^1,\ldots,Y^{\Np}) 
\cU \Omega_{\rm F}\in\cH'_{\rm phys}$
with $\phi_0 = \exp(-\frac{2\pi^2}{L}\sum_{\alpha,\beta,\gamma}
|\M_\gamma| (a_\gamma)^\alpha(a_\gamma)^\beta
 Y^\alpha Y^\beta)$ (cf.\ Eq.\ \Ref{mcM1}) 
is {\em one} groundstate of the model. One then can check that for 
any $\phi\in L^2(\R^{\Np})$ which is an eigenstate of $h_0$
and any $\psi\in \cH_{\rm Fermion}'$ which is 
a common eigenvector of all $h_p$, the state
$$
\phi(\tilde Y^1,\ldots,\tilde Y^{\Np})\RR^{\vm}\Psi = 
\phi\left(Y^1 +\f{k^1}{e^1},\ldots , Y^{\Np}  +
\f{k^{\Np}}{e^{\Np}} \right) 
\RR^{\sum_\alpha k^\alpha \mq\vn^\alpha}
\Psi\: ,    
$$
with $(\vn^\alpha)^T\cdot\vm=1$ for all $\alpha$, 
is an eigenstate of $H$ with an eigenvalue of the
form $E+c(\vm)$ where 
$$
c(\vm) = \frac{2\pi}{L}\left(\vm^2 - \sum_{j=1}^{\Np} 
(\vet_j \cdot \mq\vm)^2 \right)
$$
with $\ve_j$ the orthonormal eigenvectors of the matrix 
$\mM$ in Eq.\ \Ref{mM}. 
With that one can check that the eigenstate of $H$ which
also are invariant under the large gauge transformations 
are
\eqa
[\Psi,\phi]^{ (\theta^1,\ldots,\theta^{\Np}) }(Y) 
=  
\frac{1}{(2\pi)^{\Np/2}}
\sum_{\vk\in\Z^{\Np}} \ee{\ii k^1 
\theta^1} \cdots \ee{\ii k^{\Np} 
\theta^{\Np}}\times \nonu \times 
\phi\left(Y^1 +\f{k^1}{e^1},\ldots , Y^{\Np}  +
\f{k^{\Np}}{e^{\Np}} \right) 
\RR^{k^1 \mq\vn^1}\cdots \RR^{k^{\Np}\mq\vn^{\Np}}
\Psi   
\eqaend
where now we have $\Np$ real $\theta$ parameters.
Similarly as for $\Np=1$ the physical Hilbert space $\cH_{\rm phys}$
of the model is spanned by the states $[\Psi,\phi]^{ (0,\ldots,0) }(Y)$, 
and one needs to renormalize the inner product
of these states, i.e.\ `drop the infinite constant $\del(0)^{\Np}$',  
to get a proper inner product on $\cH_{\rm phys}$. 

\bigskip

\noindent {\bf Acknowledgement:} 
We thank M.~Luescher for interesting
discussions which prompted this work.

\end{document}